\documentclass[reqno]{amsart}

\usepackage{graphicx}

\newtheorem{lemma}{Lemma}
\newtheorem{theorem}{Theorem}
\newtheorem{proposition}{Proposition}

\newtheorem{remark}{Remark}

\newcommand{\bee}{\begin{eqnarray}}
\newcommand{\eee}{\end{eqnarray}}
\newcommand{\be}{\begin{eqnarray*}}
\newcommand{\ee}{\end{eqnarray*}}
\newcommand{\R}{{\mathbb R}}
\newcommand{\N}{{\mathbb N}}

\newcommand{\sn}{\mbox {\rm sn}}

\newcommand{\cn}{\mbox {\rm cn}}
\newcommand{\cs}{\mbox {\rm cs}}

\newcommand{\sh}{\mbox {\rm sech}}
\newcommand{\ch}{\mbox {\rm cosech}}
\newcommand{\tth}{\mbox {\rm tanh}}

\newcommand{\V}{{\mathcal V}}
\newcommand{\Hh}{{\mathcal H}}
\newcommand{\T}{{\mathcal T}}

\newcommand{\KEll}{{\mathcal K}}

\newcommand{\en}{E}
\newcommand{\En}{\Omega}

\begin{document}
 
 \title {Mathematical results for the nonlinear Winter's model}
 
 \author {Andrea Sacchetti}
  
\address {
Department of Physics, Informatics and Mathematics, University of Modena and Reggio Emilia, Modena, Italy.
}

\email {andrea.sacchetti@unimore.it}

\date {\today}

\thanks {Andrea Sacchetti is member of Gruppo Nazionale per la Fisica Matematica of Istituto Nazionale di Alta Matematica (GNFM-INdAM). \ This work is partially supported by the Next Generation EU - Prin 2022CHELC7 project "Singular Interactions and Effective Models in Mathematical Physics"  and the UniMoRe-FIM project  ``Modelli e metodi della Fisica Matematica''. \ The author is grateful to Riccardo Adami, Raffaele Carlone and Diego Noja for useful discussions.}

\begin {abstract} 

In recent years, Winter's nonlinear model has been adopted in theoretical physics as the prototype for the study of quantum resonances and the dynamics of observables in the context of  nonlinear Schr\"odinger equations. \ However, its mathematical treatment still has several important gaps. \ This article demonstrates a dispersive estimate of the evolution operator, from which the result of local well-posedeness of the solution follows; a criterion for the existence of the blow-up phenomenon is also provided. \ Finally, the phenomenon of bifurcations of stationary solutions is analysed, concluding with a conjecture on the orbital stability of some of them.

\bigskip

{\it Data availability statement.} \ All data generated or analysed during this study are included in this published article.

\bigskip

{\it Conflict of interest statement.} \ The author has no competing interests to declare that are relevant to the content of this article.
\end{abstract}




\maketitle

\section {Introduction}\label {S1}

The tunnel effect was initially studied by Gamow \cite {Ga} in order to explain
the $\alpha$-emission from a radioactive nucleus. \ In fact, in quantum mechanics, contrary to
what happens in classical mechanics, a particle can move, with a finite nonzero probability, from
a region to a different region even if these regions are separated by a potential barrier. \ In 1961 Winter proposed a simple but effective one-dimensional model \cite {Wi} for the study of the Gamow effect. \ In the Winter's model (also called single $\delta$-shell model) the potential consists of a Dirac's $\delta$ potential at $x=a > 0$, representing the internal barrier, and with a zero Dirichlet boundary condition at $x = 0$ representing the infinite barrier at the origin. \ The associated one-dimensional linear Schr\"odinger operator $\Hh_\alpha $ is formally defined on $L^2(\R^+ )$ as  
\bee
\Hh_\alpha = - \frac {\partial^2}{\partial x^2} + \V  \ \mbox { and } \ 
\V(x) = 
\left \{
\begin {array}{ll}
+\infty & \mbox { if } x < 0 \\
\alpha \delta (x-a) & \mbox { if } x \ge 0 
\end {array}
\right. \, , \ \delta_a (x)= \delta (x-a) \, ,  \label {Eq1}
\eee
where $a>0$ and $\alpha \in \R$, and where $\delta (x)$ is the Dirac's $\delta$ supported at $x=0$. \ This model has been studied in detail \cite {AgSa1,AgSa2,AC,de,W1,W2}.

New lines of research in the framework of Bose-Einstein condensates or in Optics are directed towards analysing the model obtained from Winter's model, or similar ones, by adding a non-linear term, see e.g. \cite {BSHG,BC,R1,DFFS,MK,R2,R3,R4,SacchettiAnPhys,R9,WRK}. 

The one-dimensional nonlinear Schr\"odinger equation we consider is thus given by (hereafter $\dot {\psi_t}$ means the derivative of the time-dependent wavefunction $\psi_t$ with respect to $t$):
\bee
\left \{ 
\begin {array}{l} 
i \dot \psi_t = \Hh_{\alpha} \psi_t + \eta |\psi_t |^{2\sigma } \psi_t \\
\left. \psi_t \right |_{t=0} (x) = \psi_0 (x)  
\end {array} \right. \, ,\ x \ge 0 \, , \eta \in \R \, , \ \sigma >0 \, , \label {Eq2}
\eee
where $\psi_0 \in L^2 (\R^+ )$, $\psi_0 (0)=0$, and where $\eta \in \R$ is the strength of the nonlinear perturbation and it may assume either positive (in which case we speak of \emph {defocusing} or \emph {repulsive} nonlinearity) and negative (in which case we speak of \emph {focusing} or \emph {attractive} nonlinearity) values. \ For $\sigma =1$, that is in the case of cubic nonlinearity, it is the so-called Gross-Pitaevskii equation. 

Nonlinear Schr\"odinger equations on the half-line, see e.g. \cite {ET,FIS,FHM,W} and the references therein, or on graphs, see e.g.  \cite {ABD,CFN} and the references therein, have been already considered  in a general setting but the specific case of the Winter's model needs of a peculiar analysis. \ The aim of this paper is to fill this gap. \ First of all a useful dispersive estimate for the linear evolution operator is given in \S \ref {S2}; as a result standard arguments prove the local well-posedeness of the Cauchy problem (\ref {Eq2}). \ From the analysis of the variance the blow-up occurrence is eventually discussed following the idea developed by \cite {ASa}. \ In \S \ref {S3} the problem of the calculation of the stationary states is addressed and some numerical experiments are performed showing the occurrence of a cascade of saddle point bifurcations; finally, a conjecture about the orbital stability of some of these  stationary states is stated. \ Appendix \ref {AppA} is devoted to the derivation of the time derivative of the variance.

Througout this paper, $C$ denotes a generic positive constant whose value may change from line to line.

\section {Preliminary results.} \label {S2}

In \S \ref {S2_1} we collect some preliminary results by \cite {Al} about the linear Schr\"odinger operator (\ref {Eq1}). \ Then, in \S \ref {S2_2} we give the expression of the kernel of the evolution operator; eventually, in \S \ref {S2_3} we prove a dispersive estimate of the evolution operator. 

Let $\alpha \in \R \setminus \{ 0 \}$, and let 
\bee
\psi (0)=0 \label {Eq3}
\eee
be the Dirichlet boundary condition at $x=0$ and  
\bee
\psi (a-0)=\psi (a+0) \ \mbox { and } \ \psi' (a+0) - \psi' (a-0) = \alpha \psi (a) \, , \mbox { where } \ \psi ' = \frac {\partial \psi}{\partial x}\, , \label {Eq4}
\eee
be the matching condition at $x=a$ where the Dirac's $\delta$ is supported. \ It is well known that the zero Dirichlet boundary condition at $x=0$ can be represented by a Dirac's $\delta$ at this point with positive infinite strength; then (\ref {Eq2}) is equivalent to the time-dependent Schr\"odinger equation on the whole real axis
\bee
\left \{ 
\begin {array}{l} 
i \dot \psi_t = H_{\alpha} \psi_t + \eta |\psi_t |^{2\sigma } \psi_t \\
\left. \psi_t \right |_{t=0} (x) = \psi_0 (x) \, ,\ x \in \R 
\end {array} \right. \, , \eta \in \R \, , \ \sigma >0 \, , \label {Eq5}
\eee
where 
\bee
\psi_0 \in L^2 (\R ) \ \mbox { is such that } \ \psi_0 (x)=0 \ \forall x \le 0 \label {Eq6}
\eee
and formally 
\be
H_\alpha = - \frac {\partial^2}{\partial x^2} + V  \ \mbox { and } \ 
V(x) = \beta \delta (x) + 
\alpha \delta (x-a) \ \mbox { where  }  \ \beta =+\infty \, . 
\ee
In fact, the solution to (\ref {Eq5}), when the initial wavefunction $\psi_0$ satisfies (\ref {Eq6}), is such that
\bee
\psi_t (x) =0 \, , \ \forall x \le 0 \, . 
\label {Eq7}
\eee

The linear operator $H_\alpha $ admits a self-adjoint extension (still denoted by $H_\alpha $) defined on the domain
\be
{\mathcal D}(H_\alpha ) = \left \{ \psi \in H^{2,1} (\R^+ ) \cap H^{2,2} (\R^+ \setminus \{ a\} ) \cap H_0^{2,2} (\R^+ \setminus \{ 0\} )  \ : \  \mbox { (\ref {Eq3}) and (\ref {Eq4}) hold true} \right \} \, . 
\ee

Hereafter, we omit the dependence on $\alpha$ when this fact does not cause misunderstanding: that is we simply denote $H_\alpha$ by $H$, and so on... \, . 

\subsection {Resolvent and spectrum of $H$}\label {S2_1}

Let 
\be
K_0 (x,k)= \frac {i}{2k} e^{ik|x|}\, , \ \Im k >0 \, , 
\ee
and let
\be
\Gamma (k) =
\left ( 
\begin {array}{cc}
-  \frac {i}{2k} & -  \frac {i}{2k} e^{ika} \\ 
 -   \frac {i}{2k} e^{ika} &  -   \frac {1}{\alpha}- \frac {i}{2k} 
\end {array}
\right )
\ee
with inverse matrix
\be
\Gamma ^{-1} (k) =\frac {2k}{{\mathcal G}(k)} 
\left (
\begin {array}{cc}
-2k - i\alpha  & i\alpha e^{i k a} \\
 i\alpha e^{i k a} & -i \alpha 
\end {array}
\right )\, , \ {\mathcal G}(k) = 2ik-\alpha+\alpha e^{2ika} \, . 
\ee
Then the resolvent operator is the integral operator
\bee
\left ( \left [ H -k^2 \right ]^{-1} \phi \right )(x) = \int_{\R^+} K (x,y,k) \phi (y) dy\, , \ \phi \in L^2 \, , \label {Eq8}
\eee
where 
\bee
K (x,y,k) = K_0(x-y,k)  - \frac {1}{4k^2} \sum_{j=1}^4 K_j (x,y,k) \label {Eq9}
\eee
is the kernel with
\be
K_1 (x,y,k) &=& \left [ \Gamma^{-1} (k)  \right ]_{1,1} e^{ik (|x|+|y|)}  = 2ik  e^{ik (|x|+|y|)} 
\left [ 1 - 
\frac { \alpha e^{2ika}  }
{{\mathcal G}(k)} 
\right ] \\ 
K_2 (x,y,k) &=& \left [ \Gamma^{-1} (k) ]\right ]_{1,2} e^{ik (|x|+|y-a|)} =  \frac {2ik\alpha }{{\mathcal G}(k)} e^{ik (|x|+|y-a|+a)} \\
K_3 (x,y,k) &=& \left [ \Gamma^{-1} (k)  \right ]_{2,1} e^{ik (|x-a|+|y|)} =  \frac {2ik\alpha }{{\mathcal G}(k)} e^{ik (|x-a|+|y|+a)} \\ 
K_4 (x,y,k) &=& \left [ \Gamma^{-1} (k) \right ]_{2,2} e^{ik (|x-a|+|y-a|)} = -\frac {2ik\alpha }{{\mathcal G}(k)} e^{ik (|x-a|+|y-a|)} \, . 
\ee

Concerning the spectrum of $H$ it is known that 
\bee 
\sigma_{ess} (H) = \sigma_{ac} (H) = [0,+\infty )\, , \label {Eq10}  
\eee
and the eigenvalues, if there, are given by $\en =k^2 <0$ where $k$ is a purely imaginary solution to the equation:
\bee
2ik-\alpha+\alpha e^{2ika} =0 \, , \ \Re k =0 \mbox { and } \  \Im k >0 \, , \label {Eq11}
\eee
obtained from the formula (2.1.13) by \cite {Al} for $\alpha_1 =+\infty$, $\alpha_2 =\alpha$, $y_1=0$ and $y_2 =a$ (according with the notation by \cite {Al}). 

We have that

\begin {proposition} \label {Prop1}
If $a\alpha < -1$ then the discrete spectrum of $H$ is not empty and it consists of just one negative real-valued eigenvalue
\bee
\en =-h^2 \, , \ h=  \frac {1}{2a} \left [ -a\alpha + W_0 \left ( a\alpha  e^{a \alpha } \right )\right ]  \, ; \label {Eq12}
\eee
the associated normalized eigenvector $\psi_E (x)$ is such that 
\bee
\| \psi_\en \|_{L^\infty } \le B \sinh (a) \ \mbox { and } \ \| \psi_\en \|_{L^1} \le B \frac {e^{ha} -1}{h} \, , \label {Eq13}
\eee
where
\bee
B = \left [ \frac {e^{ha} \sinh (ha) - ha}{2h} \right ]^{-1/2}\, . \label {Eq14}
\eee
If $a\alpha \ge -1$ then the discrete spectrum of $H$ is empty.
\end {proposition}

\begin {proof} Equation (\ref {Eq11}) has complex-valued solutions
\bee
k_n= \frac {i}{2a} \left [ -a \alpha +  W_n \left ( a\alpha  e^{a\alpha } \right )\right ] \, , \label {Eq15}
\eee
where $W_n (z)$ denotes the $n$-th branch of the Lambert special function, then solutions (\ref {Eq15}) are purely imaginary and such that $\Im k >0$ only if $n=0$ and $a\alpha < -1$.

Under the condition $a\alpha <-1$ we look for the normalized solution $\psi \in L^2 (\R )$ to the equation $H\psi = \en \psi$. \ The general solution to this equation satisfying the Dirichlet boundary condition at $x=0$ and the condition $\psi (x) \to 0$ as $x \to + \infty$ is given by
\be
\psi (x) = 
\left \{ 
\begin {array}{ll}
A e^{hx} & \ \mbox { if } \ x <0  \\
B \sinh (h x ) & \ \mbox { if } \ x\in (0,a) \\
C e^{-hx} & \ \mbox { if } \ x >a \, . 
\end {array}
\right.  \ ,\ h =\sqrt {-\en }\, . 
\ee
The Dirichlet boundary condition at $x=0$ implies that $A=0$, and the  matching condition at $x=a$ implies that
\be
\left \{ 
\begin {array}{lcl}
B \sinh (ha )&=& C e^{-ha}\\
h B \cosh (ha)+ \alpha B \sinh (ha) &=& -h C e^{-ha}
\end {array}
\right.
\ee
from which follows that 
\be
h=\frac {1}{2a} \left [ -a\alpha + W_0 \left ( a\alpha  e^{a \alpha } \right )\right ] >0  \, , \ \mbox { when }  k=ih; 
\ee
in fact $h$ is a solution to the equation ${\mathcal G}(ih) =0$. \ The normalized eigenvector associated to $E$ is given by
\bee
\psi_\en (x) = 
B \left \{ 
\begin {array}{ll}
0 & \ \mbox { if } \ x <0  \\
 \sinh (h x ) & \ \mbox { if } \ x\in (0,a) \\
 e^{ha} \sinh (ha) e^{-hx} & \ \mbox { if } \ x >a \, . 
\end {array}
\right. \, , \label {Eq16}
\eee
where a straightforward calculation proves (\ref {Eq14}). \ Furthermore, (\ref {Eq13}) simply follows. 
\end {proof}

\subsection {Evolution operator} \label {S2_2} The solution $\psi_t (x) \in L^2  $ to the time-dependent linear Schr\"odinger equation $i \dot \psi_t = H \psi_t$ at the instant $t$ is given by $\psi_t =e^{-itH} \psi_0$, where $e^{-itH}$ is the evolution operator associated to the self-adjoint operator $H$ and where $\psi_0 $ is the initial wavefunction satisfying (\ref {Eq6}). \ 
Let $\en $ be the eigenvalue of $H$ if $a\alpha <-1$, and let $\psi_\en$ be the associated eigenvector given by (\ref {Eq16}). \ Then
\be
e^{-i t H} \psi_0 = e^{-i t \en}  \langle \psi_\en , \psi_0 \rangle \psi_\en + e^{- i t H} P_c \psi_0
\ee
where $P_c$ is the projection operator on the eigenspace associated to the continuous spectrum of $H$ and where the expression of the evolution operator $e^{- i t H} P_c $ can be recovered from the resolvent operator. \ We observe that $\left [ P_c \psi_0 \right ] (x) =0$ for any $x \le 0$ since $P_c \psi_0 = \psi_0 - \langle \psi_E , \psi_0 \rangle \psi_E$ where $\psi_0$ and $\psi_E$ are identically zero for $x \le 0$.   

By means of the spectral theorem it turns out that the evolution operator is an integral operator 
\bee
\left [ e^{-itH} P_c \psi_0 \right ] (x) = \int_{\R} U(x,y,t) \left [ P_c \psi_0 \right ] (y) dy = \int_{\R^+} U(x,y,t) \left [ P_c \psi_0 \right ] (y) dy \label {Eq17}
\eee
where, from (\ref {Eq8}), the kernel $U(x,y,t)$ has the form
\be
U (x,y,t) &=& - \frac {i}{2\pi} \int_{\sigma_c (H)}  e^{-i\lambda t} K (x,y,k ) d\lambda \\
 &=& - \frac {i}{\pi} \int_{\R  } k e^{-ik^2 t} K (x,y,k ) dk =  U_0 (x,y,t)  + \sum_{j=1}^4 U_j (x,y,t) 
\ee
where $\lambda =k^2$ and 
\be
U_0 (x,y,t) = - \frac {i}{\pi} \int_{\R  } k e^{-ik^2 t} K_0 (x-y,k ) dk = \frac {1}{\sqrt {4\pi i t}}e^{i|x-y|^2 /4t}
\ee
and
\bee
U_j (x,y,t) 
= \frac {i}{4\pi} \int_{\R  } \frac {1}{k} e^{-ik^2 t} K_j (x,y,k ) dk\, . 
\label {Eq18}
\eee
By means of a straightforward calculation it follows that 
\bee
\sum_{j=1}^4 K_j (x,y,t)
&=& 2i k  e^{ik (|x|+|y|)}  + \frac {2i\alpha k}{{\mathcal G}(k)
} q(k,x,y) \label {Eq19}
\eee
where
\be
q(k,x,y) &=& - e^{ik (|x|+|y|+2a)}  + e^{ik (|x-a|+|y|+a)} + e^{ik (|x|+|y-a|+a)} - e^{ik (|x-a|+|y-a|)}\, .
\ee
Hence
\be
\sum_{j=1}^4 U_j (x,y,t) = \frac {i}{4\pi} \int_{\R } \frac 1k e^{-i k^2 t} \left [ 2 i k e^{ik (|x|+|y|)} + \frac {2i\alpha k}{{\mathcal G}(k)} q (k,x,y) \right ]\,. 
\ee

\begin {remark} \label {Nota1}
By means of a straightforward calculation it turns out that $U(x,y,t)=0$ if $x\cdot y<0$; thus, $\left [ e^{-itH} \psi_0 \right ] (x) =0$ for any $x \le 0$. 
\end {remark} 

Hereafter, we may assume, for argument's sake, that $x,y \ge 0$. \ In particular, 
\be
q(k,x,y) &=& 
\left \{
\begin {array}{ll}
4 e^{ik2a} \sin (kx) \sin (ky)  & \ \mbox { if } 0\le x,y \le a \\
2i e^{iky} \sin (kx) \left [ 1- e^{i2ka} \right ] &\  \mbox { if } 0\le x\le a \, , \ y >a \\
2i e^{ikx} \sin (ky) \left [ 1- e^{i2ka} \right ] &\  \mbox { if } 0\le y\le a \, , \ x >a \\
2 e^{ik(x+y)} \left [ 1- \cos (2ka) \right ] &\  \mbox { if } \ x,y >a
\end {array}
\right. \, . 
\ee
Finally, we remark that $q(k,x,y) = k^2 g(k,x,y)$ where $g(k,x,y)$ is an analytic function with respect to $k$; furthermore, $|q(k,x,y)| \le 4$ for any real-valued $k,x,y$.

\subsection {Dispersive estimate} \label {S2_3}

Now, we are going to give a typical dispersive estimate for the evolution operator.

\begin {theorem} \label {Teo1}
The following dispersive estimate holds true; let $\varphi \in L^2$, then 
\be
\| e^{-it H} P_c \varphi \|_{L^\infty} \le C t^{-1/2} \| P_c \varphi \|_{L^1} \, ,\ t >0\, , 
\ee
for some positive constant $C$ independent of $\varphi $.
\end {theorem}

\begin {remark}\label {Nota2}
If $a\alpha \ge -1$ then $P_c \varphi = \varphi$ and the the previous estimate implies that 
\bee
\| e^{-it H} \varphi \|_{L^\infty} \le C t^{-1/2} \| \varphi \|_{L^1} \label {Eq20}
\eee
for some constant $C>0$. \ If $a \alpha <-1$ then we remark that $\tilde \varphi := P_c \varphi = \varphi - \langle \psi_\en , \varphi \rangle \psi_\en$ and then 
\be
\| e^{-it H}\tilde \varphi \|_{L^\infty} &\le & C t^{-1/2} \| \tilde \varphi \|_{L^1} \le C t^{-1/2} \left [ \| \varphi \|_{L^1}  + | \langle \psi_\en , \varphi \rangle | \, \| \psi_\en\|_{L^1} \right ] \\
&\le & C \left [ 1+ \| \psi_\en \|_{L^\infty } \| \psi_\en \|_{L^1} \right ] t^{-1/2} \| \varphi \|_{L^1} \\
&\le & C t^{-1/2} \| \varphi \|_{L^1}
\ee
for some positive constant $C$. \ In conclusion, if $a \alpha <-1$ 
\be
\| e^{-it H}P_c  \varphi \|_{L^\infty} \le C t^{-1/2} \| \varphi \|_{L^1} \, , \ t >0\, , 
\ee
for some positive constant $C$ independent of $\varphi \in L^2$.
\end {remark}

\begin {remark} \label {Nota3}
From the above dispersive estimate and from the comments in Remark \ref {Nota2} we can apply the results by \cite {KT} to the operator $e^{-i t H} P_c$ obtaining thus the Strichartz estimate
\bee
\| e^{-i t H} P_c \psi \|_{L_t^p L_x^q} \le C \| \psi \|_{L^2} \label {Eq21}
\eee
for some positive constant $C$ and for any admissible pair $(p,q) \not= (2,\infty )$ such that
\be
\frac 2p +\frac 1q = \frac 12 \, , \ p,q \in [2,+\infty ] \, . 
\ee
In the case $a\alpha \ge -1$ then (\ref {Eq21}) simply becomes 
\bee
\| e^{-i t H} \psi \|_{L_t^p L_x^q} \le C \| \psi \|_{L^2} \, . \label {Eq22}
\eee
\end {remark}

\begin {proof} Let us denote by $C$ any positive constant independent of $t$, $\varphi$, $x$ and $y$; furthermore we assume that $t \ge 1$. \ From (\ref {Eq18}) and (\ref {Eq19}) it follows that 
\be
U (x,y,t) 
&=& \frac {1}{\sqrt {4\pi i t}} e^{i|x-y|^2 /4t} - \frac {1}{2\pi} \int_{\R  }     e^{ik (|x|+|y|)}  e^{-ik^2 t}  dk +\\ 
&& \ \ -\frac {\alpha }{2\pi} \int_{\R  } e^{-ik^2 t} \frac {q(k,x,y)}{{\mathcal G}(k)
}   dk\\ 
&=& \frac {1}{\sqrt {4\pi i t}}\left [ e^{i|x-y|^2 /4t} -  e^{i(|x|+|y|)^2 /4t} \right ] -\frac {\alpha }{2\pi} \int_{\R  } e^{-ik^2 t} \frac {q(k,x,t) }{{\mathcal G}(k)
}  dk
\ee
In order to estimate the integral 
\be
I:= \int_{\R  } e^{-ik^2 t} \frac {q(k,x,y)}{{\mathcal G}(k)} dk\, , \ {\mathcal G}(k) = 2 i k - \alpha \left [ 1-e^{2ika} \right ] \, ,  
\ee
we remark that $q\sim k^2$ as $k$ goes to zero, and that  $ {\mathcal G}(k) \not=0 $, for any $k \in \R \setminus \{ 0 \}$, and  ${\mathcal G}(k)=0$ at $k=0$; in particular, this zero has multiplicity one if $a\alpha \not= -1$ and it has multiplicity two when $a \alpha =-1$. \ Then, the stationary phase method suggests that $I \sim t^{-1/2}$, but a detailed analysis should be done to obtain an asymptotic behaviour uniformly with respect to $x,y \in \R^+$. \ To this end we assume, for argument's sake, that $x,y > a$; the other three cases (i.e.: $x>a$ and $0 \le y \le a$, $0\le x \le a$ and $y >0$, $0\le x,y \le a$) can be similarly treated. \ For $x,y>a$ then
\be
I =2\int_{\R  } e^{-ik^2 t+ik(x+y)} \frac {\left [ 1- \cos (2ka) \right ] }{{\mathcal G}(k)} dk \, . 
\ee
If we set
\be
w(k) = \left \{ 
\begin {array}{ll}
2i a & \ \mbox { if } \ k=0 \\
\frac {e^{2ika}-1}{k} & \mbox { if } k \not= 0 
\end {array}
\right. 
\ee
such that $w(k) \sim k^{-1}$ for $|k| \gg 1$, then
\be
{{\mathcal G}(k)} = {2ik}\left (1+ \frac {\alpha w}{2i}\right ) \ \mbox { and } \ 
\frac {1}{{\mathcal G}(k)} = \frac 1{2ik} - \frac {1}{2ik} \frac {\frac {\alpha w}{2i}}{1+\frac {\alpha w}{2i}}\, . 
\ee
Therefore
\be
I=I_a + I_b
\ee
where
\be
I_a &:=& I_a (z) = 
\int_{\R  } e^{-ik^2 t+ikz} \frac {\left [ 1- \cos (2ka) \right ] }{ik} dk \\
I_b &:=& I_b (z) =  -\int_{\R  } e^{-ik^2 t+ikz} \frac {\left [ 1- \cos (2ka) \right ] }{ik} \frac {\frac {\alpha w}{2i}}{1+ \frac {\alpha w}{2i}} dk
\ee
where we denote 
\be
z=x+y >2a \, . 
\ee

\begin {lemma} \label {Lemma1} We have that 
\be
I_a (z) = \frac {1+i}{8}\pi (-1)^{3/4} \sqrt {2}  \left [ -2\mbox {\rm erf} \left (\tilde A z/\sqrt {t} \right ) - \mbox {\rm erf} \left (\tilde B/\sqrt {t} -\tilde A z/\sqrt {t} \right ) +  \mbox {\rm erf} \left (\tilde B/\sqrt {t} +\tilde A z/\sqrt {t}  \right ) \right ]
\ee
where
\be
\tilde A= \frac 14 (1-i) \sqrt 2 \ \mbox { and } \ \tilde B = 2a \tilde A \, ,
\ee
and in particular
\bee
|I_a (z)| \le \frac {a}{4\sqrt {\pi}}t^{-1/2}\, , \ \forall z \in \R \, . \label {Eq23}
\eee
\end {lemma}

\begin {proof} Indeed, the derivative $I_a'$ of $I_a$ with respect to $z$ is given by 
\be
I_a' (z) &=& \int_{\R  } e^{-ik^2 t+ikz}  {\left [ 1- \cos (2ka) \right ] }dk \\
&=& \int_{\R  } e^{-ik^2 t+ikz} dk - \frac 12 \int_{\R  } e^{-ik^2 t+ikz+2kai} dk - \frac1 2 \int_{\R  } e^{-ik^2 t+ikz-2kai} dk \\
&=& \frac {\sqrt {\pi}}{\sqrt {4it}} e^{i z^2/4t}- \frac 12 \frac {\sqrt {\pi}}{\sqrt {4it}} e^{i(2a+z)^2/4t} -\frac 12 \frac {\sqrt {\pi}}{\sqrt {4it}} e^{i(-2a+z)^2/4t}\, . 
\ee
Thus, we have a differential equation for $I_a (z)$ with condition $\lim_{z\to + \infty } I_a (z) =0$ which solution is simply given by 
\be
I_a = -\int_z^{+\infty} I_a' (u) du \, . 
\ee
Then Lemma \ref {Lemma1} follows since
\be
- \int_z^{+\infty } \frac {\sqrt {\pi}}{\sqrt {4 i t}}e^{iu^2/4t} du = -\frac 14  (1+i)\sqrt {2} \pi (-1)^{3/4} \left [ -1 + \mbox {erf} \left ( \frac {(1-i) \sqrt {2} z}{4\sqrt {t}} \right ) \right ] \, . 
\ee
Recalling that
\be
\mbox {erf} (z)= \frac {2}{\sqrt {\pi}} \int_0^z e^{-w^2 } dw 
\ee
then a straightforward calculation gives that 
\be
I_a (z) 
&=& \frac {1+i}{\sqrt {8\pi}} (-1)^{3/4}  \int_{ \tilde A z/\sqrt {t}-2a\tilde A/\sqrt{t} }^{\tilde A z/\sqrt {t} + 2a\tilde A/\sqrt{t}}e^{-w^2} dw \\
&=&\frac {2i}{\sqrt {32\pi}} (-1)^{3/4} \int_{ \frac 14 \sqrt {2} \left [ z/\sqrt {t}-2a/\sqrt{t}\right ] }^{ \frac 14 \sqrt {2} \left [ z/\sqrt {t}+2a/\sqrt{t}\right ] }e^{-\frac {i}{2}\tau^2} d\tau
\ee
where we set $w(1-i)=\tau$. \ Hence (\ref {Eq23}) follows. 
\end {proof}

Now, we estimate the second integral 
\be
I_b (z) = - \int_{\R  }  e^{-ik^2 t + i k z} Q(k) dk 
\ee
where 
\be
Q(k) = \frac {\left [ 1 -\cos (2 k a) \right ]}{ik} \frac {\frac {\alpha w(k)}{2i}}{1+\frac {\alpha w(k)}{2i}} 
\, . 
\ee
In particular, 
\be
Q(k) 
= - i \alpha \frac {\sin^2 (ka)}{k^2} \frac {\sin (ka)}{e^{-ika} + \frac {\alpha}{k} \sin (ka)}\ \mbox { for } \ k \not= 0 \, . 
\ee
We also remark that 
\bee
Q(k) \sim 
\left \{
\begin {array}{ll}
- \frac {2ia^3\alpha}{1+a\alpha } k & \mbox { if } \ 1+a\alpha \not= 0 \\
-2a - \frac 23 i a^2 k  & \mbox { if } \ 1+a\alpha = 0
\end {array}
\right. \ \mbox { as } \ |k| \ll 1 \, . \label {Eq24}
\eee
We have that

\begin {lemma} \label {Lemma2} The following estimates hold true 
\bee
|Q(k) | \le \min \left [ |\alpha | a^2 , \frac {|\alpha |}{k^2} \right ] \, , \ \forall k \in \R \setminus \{0 \} \, , \label {Eq25}
\eee
and
\bee
|Q' (k) | \le C \, , \ \forall k \in \R \, , \ \mbox { and } \ |Q' (k) | \le \frac {C}{k^2} \, , \ \forall |k| \ge \pi /a \, , \ \mbox { where } \ Q' = \frac {d Q}{d k} \, . \label {Eq26}
\eee
Finally
\bee
|Q'' (k)|\le C \, , \ \forall k \in \R \,  , \ \mbox { where } \ Q'' = \frac {d^2 Q}{d k^2} \, .
\label {Eq27}
\eee
\end {lemma}

\begin {proof}
Since $|\sin (ka ) |\le \min [1,|k|a] $ and 
\be
\left | e^{-ika} + \frac {\alpha}{k} \sin (ka) \right | \ge | \sin (ka )|  
\ee
then (\ref {Eq25}) follows. \ Concerning the derivative
\be
Q' (k) = \alpha \frac {\sin^2 (ka)}{k^2} \frac {q_1 (k)}{p_1 (k)} 
\ee
where
\be
q_1 (k) = \left [ 2i k \sin (ka)  - 3i k^2 a \cos (ka) + a k^2 \sin (ka ) \right ] e^{-ika} + i \alpha \sin (ka) \left [ \sin (ka) - 2 k a \cos (ka ) \right ]
\ee
and
\be
p_1 (k) = f^2 (k) \ \mbox { where } \ f(k) =  k e^{-ika} + \alpha \sin (ka) 
\ee
are such that
\be
q_1 (k) \sim -i a (1+a\alpha ) k^2 + i a^3 \left ( \frac 23 + a\alpha \right ) k^4 + O(k^5) \, ,\ \mbox { for } |k|\ll 1 \, , 
\ee
and 
\be
p_1 (k) \sim (1+a\alpha )^2 k^2 - 2 i a (1+a \alpha ) k^3 -a^2 \left ( \frac 13 a^2 \alpha^2 + \frac 43 a \alpha +2 \right ) k^4 + O(k^5) \, ,\ \mbox { for } |k|\ll 1 \, .
\ee
Hence, there exists $k^\star >0$ such that 
\bee
\left | \frac {q_1 (k)}{p_1 (k)} \right | \le C \, , \ \forall k \in [-k^\star ,+k^\star ] \label {Eq28}
\eee
for some $C>0$. \ We consider now the case $k \notin [-k^\star ,+k^\star ]$. \ To this end we observe that
\be
|q_1 (k) | \le C k^2 \, ,\ \mbox { for any } k  \, .
\ee
For what concerns $p_1 (k) = f^2(k)$ we have that  
\be
 f(k) = \left [ k \cos (ka) + \alpha \sin (ka) \right ] - i k \sin (ka) \, . 
\ee
Let $\delta \in (0,1)$ be suitably chosen and let $k \notin [-k^\star ,+ k^\star ]$ be such that
$ |\cos (ka) |<\delta $; in this case
\be
 |f(k) |\ge |k| \sqrt {1-\delta^2}\, . 
 \ee
 Let us consider now the case where $k \notin [-k^\star ,+ k^\star ]$ is such that
$\delta \le  |\cos (ka) | $; in this case 
\be
|f(k)| \ge 
|k| \, |\cos (ka) | - |\alpha |\, |\sin (ka)| 
\ge \delta |k| - \alpha \sqrt {1-\delta^2} 
\ge \delta |k| 
\left [ 1- 
\frac 
{\alpha \sqrt {1-\delta^2}}
{\delta k^\star } \right ]\, . 
\ee
For $\alpha$ and $k^\star$ fixed we choose $\delta$ such that $\frac 
{\alpha \sqrt {1-\delta^2}}
{\delta k^\star }=\frac 12$. \ Therefore, we can conclude that 
\be
|f(k)| \ge C |k| \, , \ \mbox { that is } |p_1 (k) | \ge C k^2 \, , \ \forall k \notin [-k^\star ,+k^\star]\,.
\ee
From this fact and from (\ref {Eq28}) then $|q_1/p_1 | \le C$ for any $k \in \R$. \ Hence, both inequalities in (\ref {Eq26}) follow.

Finally, in order to prove (\ref {Eq27}) we simply remark that (by $p_1'$ and $q_1'$ we rispectively  denote the derivatives of $p_1$ and $q_1$ with respect to $k$)
\be
Q'' (k) = 2  \alpha \frac {\sin (ka) [ ka\cos (ka) - \sin (ka) ]}{k^3 } \frac {q_1 (k)}{p_1 (k)} + \alpha \frac {\sin^2 (ka)}{k^2 } \left [ \frac {q_1'p_1 -q_1 p_1'}{p_1^2} \right ] 
\ee
By means of a straightforward calculation it follows that
\be
| ka\cos (ka) - \sin (ka) | \le
\left \{ 
\begin {array}{l}
C |k|^3 \, , \ \mbox { if } \ |k| \le \tilde k  \\ 
C |k| \, , \ \mbox { if } \ |k| \ge \tilde k  \, . 
\end {array}
\right. \, ,
\ee
for some $\tilde k >0$.
Furthermore
\be
p_1(k) q_1'(k)-p_1' (k) q_1 (k) & \sim & 2a^2 (a\alpha +1)^2 k^4 + i a^3 \left ( - \frac 83 +2 a \alpha +2a^2 \alpha^2 + \frac 43 a^3 \alpha^3 \right ) k^5 + O(k^6) 
\ee
for $|k| \ll 1$, and 
\be
|p_1(k) q_1'(k)-p_1' (k) q_1 (k)| \le C k^4 \, , \ \mbox { for } \ |k| \gg 1 \, .
\ee
Hence $|Q'' (k)|\le C$ for any $k\in \R$.
\end {proof}

Then
\be
I_b (z) =-2 e^{iz^2/4t} \int_{\R  } e^{-iu^2 t} P(u)du = I_b^+ (z) + I_b^- (z)
\ee
where we set $k=u+z/2t$ and $P(u) =  Q(u+z/2t) $ and where
\be
I_b^\pm (z) = \mp 2 e^{iz^2/4t} \int_{0 }^{\pm \infty  } e^{-iu^2 t} P(u)du \, . 
\ee
Let us represent $P(u)$ in the form
\be
P(u) = P(0)+u P_1 (u) \ \mbox { where } \ P_1 (u) = 
\left \{ 
\begin {array}{l}
\frac {P(u)-P(0)}{u}\, , \ \mbox { if } u \not= 0 \\
\frac {dP(0)}{du} \, , \ \mbox { if } u=0 
\end {array}
\right. \, . 
\ee 
Then
\be
I_b^\pm = I_{b,0}^\pm + I_{b,1}^\pm  
\ee
where
\be
I_{b,0}^\pm 
= \mp 2 e^{iz^2/4t} \int_{0}^{\pm \infty } e^{-iu^2 t} P(0)du 
\ee
and
\be  
I_{b,1}^\pm = \mp 2 e^{iz^2/4t} \int_{0 }^{\pm \infty  } e^{-iu^2 t} u P_1(u)du \, . 
\ee
Concerning the first integral it turns out that 
\be
I_{b,0}^+ &=& 2 P(0) e^{iz^2/4t} \int_0^{+ \infty} e^{-iu^2 t} du = O(t^{-1/2} ) 
\ee
since $|P(0)|=|Q (z/2t)| \le C$ and
\be 
\int_0^{+ \infty} e^{-iu^2 t} du  
= t^{-1/2} \int_0^{+ \infty} e^{-iv^2} dv =  t^{-1/2} \left ( - \frac 12 \right ) (-1)^{3/4} \sqrt {\pi} \, . 
\ee
Concerning the second integral it turns out that 
\be
I_{b,1}^+ &=&  -2 e^{iz^2/4t} \int_{0 }^{+ \infty  } u e^{-iu^2 t} P_1(u)du \\ &=&  -2 e^{iz^2/4t} \left \{ \left [ \frac i{2t}  e^{-iu^2 t} P_1(u) \right ]_0^{+\infty} -   \frac i{2t} \int_{0 }^{+\infty  }  e^{-iu^2 t} \frac {dP_1(u)}{du} du \right \} \, . 
\ee
Therefore
\be
|I_{b,1}^+ | \le \frac 1t \left [ |P_1(0)| + |P_1 (+\infty )| + \int_{0 }^{+\infty  } \left | \frac {dP_1(u)}{du} \right | du \right ]
\ee
where $P_1 (0) = \frac {dP(0)}{du}  =\left. \frac {dQ(u)}{du} \right |_{u=z/2t}$ and $P_1 (+\infty )=0$. \ In order to estimate the integral
\be
\int_0^{+\infty} \left | \frac {dP_1(u)}{du} \right | du 
\ee
let us remark that
\be
\frac {dP_1(u)}{du}  = \frac {u\frac {dP(u)}{du}  -P(u)+P(0)}{u^2} = \frac {1}{u^2} \left [ u \frac {dQ(u+z/2t)}{du}  - Q(u+z/2t) + Q(z/2t) \right ]
\ee
and let us set
\be
A= \left \{ \left [ - \frac {z}{2t} - \frac {\pi}{a}  , - \frac {z}{2t} + \frac {\pi }{a} \right ] \cup \left [0, k^\star \right ] \right \} \cap [0,+\infty ) \ \mbox { and } \ A^C = [0,+\infty ) \setminus A \, .
\ee
Then, if we prove that 
\be
\int_A \left | \frac {dP_1(u)}{du} \right |  du \le C \ \mbox { and } \ \int_{A^C} \left | \frac {dP_1(u)}{du} \right | du \le C
\ee
then the result follows. \ Indeed, from (\ref {Eq25}) and (\ref {Eq26}) it follows that 
\be
\int_{A^C} \left | \frac {dP_1(u)}{du} \right | du \le \int_{A^C} \frac {1}{u^2} \left [ \frac {C|u|}{(u+z/2t)^2}+ C\right ] du \le C
\ee
since
\be
\int 
\frac {1}{u(u+\gamma )^2} du = \frac {1}{\gamma^2} \ln \left ( \frac {u}{u+\gamma} \right ) + \frac {1}{\gamma (u+\gamma) } \le C \, , \  \forall \gamma \not= 0 \, . 
\ee
Concerning the other integral we simply remark that there exists a function $\xi (u)$ such that 
\be
\frac {dP_1(u)}{du}  =- \frac 12 \left. \frac {d^2 P(v)}{dv^2} \right |_{v=  \xi (u) } \, , \ |\xi (u)| \le |u|\, , 
\ee
where $\frac {d^2P (u)}{du^2}  = \frac {d^2Q (u+z/2t)}{du^2}  $. \ Since $ \left | \frac {d^2Q (u}{du^2} \right |\le C$ then 
\be
\int_{A} \left | \frac {dP_1(u)}{du}  \right | du \le C \, . 
\ee
In conclusion, we have proved that 
 \be
 | I^+_{b,1} (z)| \le \frac Ct 
 \ee
 for some constant $C$ independent of $z$, and thus $I_b ^+ (z) \sim t^{-1/2}$ uniformly with respect to $z\in \R$. \ Since the same result similarly follows for $I_b^-$ then the Theorem is proved when $x,y > a$. \ The other three cases ($x,y\le a $, $x\le a$ and $y>a$, and $x>a$ and $y\le a$) can be similarly treated. \ Hence, the proof of the Theorem is completed.  
\end {proof}

\section {Well-posedeness of the NLS} \label {S3} In this section we consider the well-posedeness problem for the nonlinear Schr\"odinger equation 
\bee
\left \{
\begin {array}{l}
i\dot \psi_t = H_\alpha \psi_t + \eta |\psi_t |^{2\sigma } \psi_t  \\
\left. \psi_t \right |_{t=0} = \psi_0 
\end {array}
\right.  \, , \ \psi_t \in L^2 (\R ) \, , \ \| \psi_0 \| = 1\, , \ \sigma >0 \, ,  \label {Eq29}
\eee
where $\psi_0$ satisfies condition (\ref {Eq6}). \ For argument's sake we assume that $\alpha >0$; in such a case the self-adjoint operator $H_\alpha \ge 0$ and it has no bound states. 

\begin {remark} \label {Nota4} Condition $H_\alpha \ge 0$ enable us to make use of some well know results, e.g. Theorem 3.7.1 by \cite {C}. \ In fact, if $\alpha <0$ then one can treat the problem of the existence of the solution replacing $H_\alpha$ by $H_\alpha + \gamma$ for some $\gamma >0$ such that $H_\alpha + \gamma \ge 0$, as done by \cite {ASa}. \ However, we don't dwell here on such a detail.
\end {remark}  

As in the previous Section let us omit the dependence on $\alpha$ when this fact does not cause misunderstanding. 

From  Theorem 3.7.1 by \cite {C} and by the  Strichartz  estimate (\ref {Eq22}) then a unique solution $\psi_t$ to (\ref {Eq29}) locally exists in $H^1$ for any $t \in [0,T_{max} )$ for some $0 < T_{max} \le +\infty $, it satisfies the boundary conditions (\ref {Eq3}) and (\ref {Eq4}) and the conservation of the norm and of the energy. \ That is 
\be
\| \psi_t \| = \| \psi_0 \| \ \mbox { and } \ {\mathcal E} (\psi_t ) = {\mathcal E} (\psi_0 )\, , \ \mbox { for any } \ t \in [0,T_{max} )\, , 
\ee
where
\be
{\mathcal E} (f ) = \langle f , H f \rangle +   \frac {\eta}{\sigma +1} \| f \|_{L^{2(\sigma +1)}}^{2(\sigma +1)} =  \left \| \frac {\partial f}{\partial x} \right \|^2 + \alpha | f (a) |^2 +  \frac {\eta}{\sigma +1}  \| f \|_{L^{2(\sigma +1)}}^{2(\sigma +1)} \, . 
\ee
Furthermore, the blow-up alternative states that 
\bee
T_{max}=+\infty \ \mbox { or } \ \lim_{t \to T_{max}-0} \| \psi_t \|_{H^1} = +\infty \, . \label {Eq30}
\eee

\begin {remark} \label {Nota5} The Dirichlet boundary condition (\ref {Eq3}) at $x=0$ basically splits the real axis in the two domains $(-\infty ,0)$ and $(0,+\infty )$ and the values of the wavefunction $\psi_t (x)$ in one of these two intervals don't influence its values on the oher interval. \ Thus, if $\psi_0 (x)=0 $ for any $ x \le 0$ then $\psi_t (x)=0$ for any $x \le $ and any $t <T_{max}$. \ This quite obvious fact can be also understood by writing the solution $\psi_t(x)$ to (\ref {Eq29}) in the Duhamel's form
\be
\psi_t = F(\psi_t) :=e^{-i t H} \left [ \psi_0 + \eta \int_0^t e^{i s H}  |\psi_s |^{2\sigma } \psi_s ds \right ] \, .
\ee 
Let $\psi^0_t (x) =\psi_0 (x)$ for any $t$, and let 
\be
\psi_t^{(n+1)} = F(\psi_t^{(n)} ) \, , \ n \ge 0 \, .
\ee
Since $\psi_0 (x)=0 $ for any $x \le 0$, then $\psi^{(n)}_t (x)=0$ for any $x \le 0$ and any $n \ge 1$, too. \ 
Therefore, the fixed point argument, that usually proves that $\psi_t^{(n)}  \to \psi_t$, implies that $\psi_t (x)=0$ for any $x\le 0$ and any $t \in [0,T_{max})$.
\end {remark}

From the conservation of the energy and from (\ref {Eq30}) then it follows that 

\begin {theorem} \label {Teo2}
If one of the following conditions

\begin {itemize}

\item [i.] $\eta \ge 0$;

\item [ii.] $\eta <0$ and $\sigma <2$;

\item [iii.] $c_1 < \eta <0$ and $\sigma =2$, where $c_1<0$ is a given constant independent of $\psi_0$;

\item [iv.]  $c_2 < \eta <0$ and $\sigma >2$, where $c_2 <0$ is a given constant dependent on $\sigma  $ and $\psi_0$;

\end {itemize}

is satisfied then there is no blow-up and the solution $\psi_t$ exists for any $t \ge 0$.
\end {theorem}

\begin {proof}
The proof is a simple adaptation of that given by \cite {ASa} in the case of the single delta, and it is therefore omitted.
\end {proof}

In order to state sufficient conditions for blow up let us define 
\be
I_q (t):= \int_{0}^{+\infty} (x-q)^2 |\psi_t (x)|^2 dx = \langle \psi_t , (x-q)^2 \psi_t \rangle \, \ \mbox { since } \ \psi_t(x)=0 \, \ \forall x \in (-\infty, 0] \, , 
\ee
be the \emph {moment of inertia} with respect to the point $q$, where $q>0$ is a given point of the real axis. \ Let $\psi_t$ be the solution to (\ref {Eq29}) then we have that the derivative of $I_q$ with respect to the time is such that (see Appendix \ref {AppA} for details)
\bee
\dot I_q = 4 \Im  \left \langle  \psi_t ,(x-q) 
\frac {\partial \psi_t}{\partial x}  \right \rangle \label  {Eq31}
\eee
and  
\bee
4 \Re \left \langle  \psi_t ,(x-q) 
\frac {\partial \psi_t}{\partial x}  \right \rangle + 2 \| \psi_t \|^2 = 0\, . \label  {Eq32}
\eee
Furthermore, a straightforward calculation gives that the second derivative is such that
\bee
\ddot I_q =  8  {\mathcal E}_0  - 4 \alpha | \psi_t (a) |^2 + \frac {4\eta (\sigma -2)}{\sigma +1} \| \psi_t \|_{L^{2(\sigma +1)}}^{2(\sigma +1)} -4\T \, . \label {Eq33}
\eee
where $\T= q \left | \frac {\partial \psi_t (0+0)}{\partial x}  \right |^2  -(a-q) \left [\left | \frac {\partial \psi_t (a+0)}{\partial x} \right |^2 - \left | \frac {\partial \psi_t (a-0)}{\partial x} \right |^2 \right ]$.

\begin {remark} \label {Nota6}
Since (\ref {Eq33}) holds true for any $q \in \R$ then for $q=a$ it takes the simplest form
\be
\ddot I_a =  8  {\mathcal E}_0  - 4 \alpha | \psi_t (a) |^2 + 4\eta \frac { (\sigma -2)}{\sigma +1} \| \psi_t \|_{L^{2(\sigma +1)}}^{2(\sigma +1)} - 4 a  \left | \frac {\partial \psi_t (0+0)}{\partial x}  \right |^2       \, . 
\ee
In particular, since $a>0$, it follows that
\bee
\ddot I_a \le  8  {\mathcal E}_0  - 4 \alpha | \psi_t (a) |^2 + 4\eta \frac { (\sigma -2)}{\sigma +1} \| \psi_t \|_{L^{2(\sigma +1)}}^{2(\sigma +1)}        \, . \label {Eq34}
\eee
\end {remark}

Therefore, we fall in the same situation as in Theorem 3 by \cite {ASa} and thus the same blow-up criterion works; it reads as follows. 

\begin {theorem} \label {Teo3}
If ${\mathcal E} (\psi_0 ) <0$, $\sigma >2$ and $\eta < \eta_c$, for some $\eta_c <0$, then blow up in the future occurs in finite time, i.e.: $T_{max}<+\infty$.
\end {theorem}

\section {Stationary states of the NLS} \label {S4}

\subsection {Preliminary results} \label {S4_1} We look for stationary solutions to the equation (\ref {Eq5}); that is $\psi_t (x) = e^{-i\En t} \psi (x)$ where ${\En}$ is real-valued and $\psi (x)$ is a solution to the time-independent NLS
\bee
H_\alpha \psi + \eta |\psi |^{2\sigma } \psi = {\En} \psi  \,  , \psi \in L^2  \, ,\ \| \psi \| =1\, , \ \alpha \in \R \, , \label {Eq35}
\eee
such that 
\be
\psi (x)=0 \, , \ \forall x \in (-\infty ,0]\, . 
\ee

First of all we prove that if a solution $\psi $ to (\ref {Eq35}) there exists then $\psi $ is, up to a constant phase factor, a real-valued function. 

\begin {proposition} \label {Prop2}
Let $\psi \in L^2 $ be a solution to the nonlinear equation  (\ref {Eq35}), where ${\En}$ and $\eta$ are real-valued, satisfying conditions (\ref {Eq3}) and (\ref {Eq4}). \ Let $\sigma \in \N$ be a positive integer number. \ Then $\psi (x)$ is, up to a constant phase factor, a real-valued function.
\end {proposition}

\begin {proof}
By means of a simple calculation it follows that  
\be
W(x):= \left [ \frac {\partial  \psi}{\partial x} \bar \psi - \psi \frac {\partial  \bar \psi}{\partial x} \right ] = 
\left \{
\begin {array}{ll}
c_0 & \mbox { if } x \in (-\infty ,0) \\
c_1 & \mbox { if } x \in (0,a) \\ 
c_2 & \mbox { if } x \in (a,+\infty ) 
\end {array}
\right. 
\ee
is a piece-wise constant function where $c_0=c_1 =0$ since $\psi (0)=0$. \ Let 
\be
W_\pm = W(a\pm 0) :=\lim_{x\to a^\pm} W(x)
\ee
be the right (+) and left (-) hand side limit of $W(x)$ at $x=a$; then conditions (\ref {Eq4}) imply that 
\be
W_+= W_-\, . 
\ee
Hence, $c_2=0$, too. \ Now, if we set
\be
\psi (x) = 
\left \{ 
\begin {array}{ll}
\psi_0 (x) := \phi_0 (x) e^{i\theta_0 (x)} & \mbox { if } x \in (-\infty ,0) \\
\psi_1 (x) :=\phi_1 (x) e^{i\theta_1 (x)} & \mbox { if } x \in (0,a) \\
\psi_2 (x) :=\phi_2 (x) e^{i\theta_2 (x)} & \mbox { if } x \in (a,+\infty)
\end {array}
\right.
\ee
where $\phi_{j}(x)  $ and $\theta_{j} (x)$, $j=0,1,2$, are real-valued functions, then equation $\frac {\partial  \psi_j}{\partial x} \bar \psi_j - \psi_j \frac {\partial \bar \psi_j}{\partial x} =0$ implies that $\phi_j =0$ or $\theta_j$ is a constant function; in the latter case then the matching conditions at $x=0$ and at $x=a$ imply that $\theta_2 - \theta_1 = 2n \pi $ and $\theta_1 - \theta_0 = 2m \pi $, for some integer numbers $n$ and $m$. 
\end {proof}

\begin {remark} \label {Nota7}
From Proposition \ref {Prop2} it follows that when  $\eta$ and $\En$ are real-valued and $\sigma \in \N$ then equation (\ref {Eq35}) takes the form
\bee
H \psi + \eta \psi^{2\sigma +1} = {\En} \psi  \, , \ \| \psi \| =1 \, . \label {Eq36}
\eee
Eventually, if we set $\psi ={|\eta |}^{-1/2\sigma } \varphi$, then it becomes
\bee
H \varphi + g \varphi^{2\sigma +1} = {\En} \varphi  \, , \  g=\pm 1 \, , \ \| \varphi \| =\mu 
\ \mbox { where } \ 
\mu :={|\eta |}^{1/2\sigma } \, . \label {Eq37}
\eee
\end {remark}

Now, we look for real-valued stationary solutions for any $\alpha \in \R$ and where $\sigma =1$, that is we focus our attention to the Gross-Pitaevskii equation with cubic nonlinearity.

\subsection {Stationary states for focusing NLS with cubic nonlinearity} \label {S4_2}

In this section we restrict our attention to the focusing nonlinearity case where $g=-1$. \ It is well known \cite {D} that the general real-valued solution to the equation
\bee
-\frac {d^2 \varphi}{dx^2} - \varphi^3 = {\En} \varphi \,  , \label {Eq38}
\eee
may be written in the form
\be
\varphi (x) = C \cn \left ( \lambda (x-x_0 ) , p \right ) \, , \ p \in [0,1]\, , 
\ee
where 
\be
p^2 = \frac {\lambda^2-{\En}}{2\lambda^2} \ \mbox { and } \ C^2 = {\lambda^2-{\En}}=  {2p^2 \lambda^2} \, ,
\ee
for some $C,\lambda \in \R$. \ Recalling that 
\be
\lim_{p \to 1^-} \cn (u,p) = \sh (u)
\ee
and since we look for a real-valued solution $\varphi (x)$  to (\ref {Eq38}) such that $\varphi (x)=0$, for any $x\le 0$, and $\varphi (x) \to 0$ as $x\to + \infty$ then such a solution there exists only when ${\En}<0$ and it has the form 
\be
\varphi (x) = 
\left \{
\begin {array}{ll}
C \cn \left ( \lambda (x-x_0 ) , p \right ) & ,\ x \in (0,a) \ \mbox { where } p^2 = \frac {\lambda^2 - \Omega}{2\lambda^2} \ \mbox { and } \ C^2 = {\lambda^2 - \Omega} \\ 
C' {\sh} \left ( \lambda' (x-x_0' ) \right ) & ,\ a< x \ \mbox { where } {\lambda'}^2 =-{\En} \ \mbox { and } \ {C'}^2 =  - {2{\En}} 
\end {array}
\right. \, .
\ee
Since the two functions $\cn (x,p)$ and $\tth (x)$ are both even functions with respect to $x$ then we may assume, for argument's sake and without loosing in generality, that $\lambda >0$ and $\lambda ' >0$.

The Dirichlet boundary condition $\varphi (0)=0$ at $x=0$ implies that $\lambda x_0 ={\KEll}(p)$ is a zero of the Jacobi elliptic function $\cn (x,p)$. \ The matching condition (\ref {Eq4}) at $x=a$ implies that 
 for $g=-1$ the real-valued parameters $C$, $C'$, $\lambda$, $\lambda'$, $ x_0$, $ x_0'$  and $ p$ must satisfy to the following set of conditions
\bee
\left \{
\begin {array}{l}
\lambda x_0 ={\KEll}(p)\\
C' {\sh} \left ( \lambda' (a-x_0' ) \right ) - C \cn \left ( \lambda (a-x_0) , p \right ) =0 \\
C' {\sh} (\lambda' (a-x_0')) \left [ \lambda' {\tth} (\lambda' (a-x_0')) + \alpha \right ] - C \lambda {\sn}(\lambda (a-x_0),p) \mbox {dn}(\lambda (a-x_0),p)  = 0 \\
C^2 = 2 p^2 \lambda^2 \\
 {C'}^2 = {2{\lambda'}^2} \\
 (1-2p^2) \lambda^2 = -{\lambda '}^2 =\Omega
\end {array}
\right. \label {Eq39}
\eee
with the constrain $2p^2 -1 \ge 0$, that is $p \in [1/\sqrt {2},1]$.

Since ${\tth} (x) =\pm \sqrt {1-{{\sh}}^2 (x)}$ then by reduction we can reduce the system of equations (\ref {Eq39}) as follows (the proof of the Lemma is just a straightforward calculation we omit).

\begin {lemma} [Effective equation for stationary solutions - attractive nonlinearity] \label {Lemma3} 
Let 
\be
w(p):=\frac {1}{\sqrt {2p^2-1}}
\ee
and let 
\be
H_\ell^f (p,\lambda ') &:=& 
\cn \left [ {\lambda'}{w(p)} a-{\KEll}(p) , p \right ] \left \{ \lambda'  (-1)^\ell  \sqrt {1-\frac {p^2}{2p^2-1}{\cn }^2 \left [ {\lambda'}{w(p)} a-{\KEll}(p) , p \right ]} + \alpha \right \} + \\ && \ \ - {\lambda'}{w(p)} {\sn}\left [{\lambda'}{w(p)} a-{\KEll}(p),p \right ] \mbox {\rm dn}\left [ {\lambda'}{w(p)} a-{\KEll}(p),p \right ] \, , \ \ \ell =1,2 \, .
\ee
If $p$, $x_0$, $x_0'$, $\lambda$, $\lambda'$ $C$ and $C'$ is a solution to (\ref {Eq39}) then the following equation must be satisfied
\bee
H_\ell^f (p,\lambda' )=0 \, ,\label {Eq40}
\eee 
for $\ell =1$ or $\ell =2$. \ The norm of the solution is given by:
\be
\mu^2 = \| \varphi \|^2 
= C^2 \int_0^a \left [ \cn \left ( \lambda (x-x_0 ) , p \right ) \right ]^2 dx + {C'}^2 \int_a^{+\infty} \left [ {\sh} \left ( \lambda' (x-x_0' ) \right ) \right ]^2 
dx  \, , \ 
\ee
and 
\be
\eta  =-\mu^2 \, .
\ee
\end {lemma}

\subsubsection {Numerical experiments.} 
\begin{figure}
\centering
\includegraphics[width=4.25in]{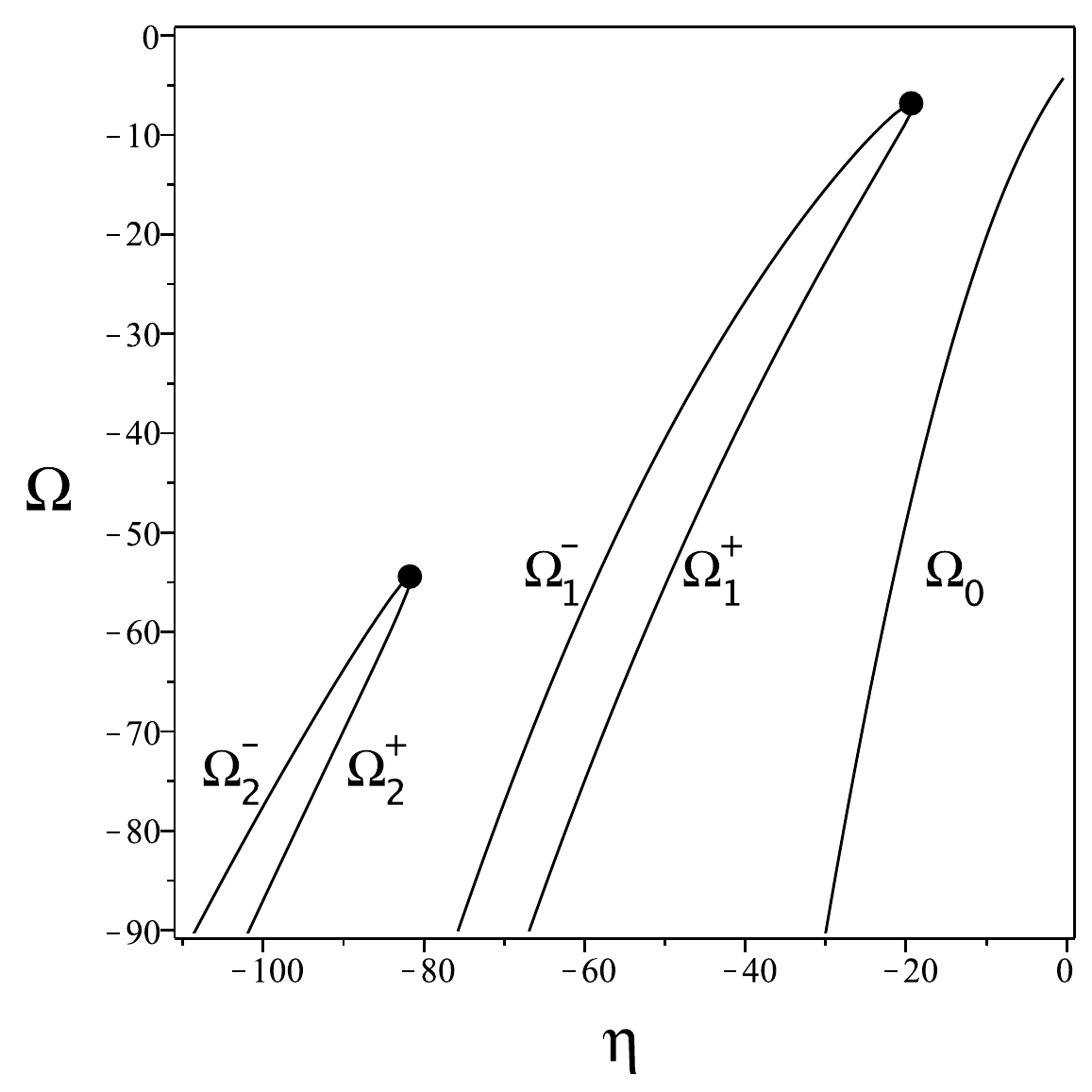}
\caption{For $a=1$ and $\alpha =-4$ then for  $\eta \in [-110,0]$  two branches $\Omega_1^\pm (\eta )$ and $\Omega_2^\pm (\eta )$ of stationary solutions raise at the bifurcation points corresponding to the two circles. \ The stationary solution obtained for the linear problem admits the continuation $\Omega_0(\eta )$.}
\label {Fig1}
\end{figure}

Here a consider the numerical experiment where 
\be
a=1 \ \mbox { and } \ \alpha =-4 \, .
\ee
When $\eta =0$ the linear Schr\"odinger operator has just one negative eigenvalue given by
\be
\Omega_0 =- \left [ \frac {1}{2a} W \left ( a\alpha e^{a\alpha } \right ) - \frac 12 \alpha \right ]^2 =-3.84(3) \, ,
\ee
where $W(x)$ is the Lambert's special function. \ For $\eta <0$ we still have one stationary solution as the continuation of the one obtained for $\eta =0$; furthermore, a sequence of saddle point bifurcation points $(\eta_n , \Omega_n )$, $n=1,2,\ldots $, occur for increasing values of $|\eta |$ and these bifurcation points are associated to two branches $\Omega_n^{\pm} (\eta )$, $\eta <\eta_n$, of stationary solutions (see Figure \ref {Fig1}). \ The bifurcation points are the solutions to the following systems:
\be
\left \{ 
\begin {array}{lcl}
H_\ell^f (p,\lambda ') &=& 0 \\
\frac {\partial H_\ell^f (p, \lambda ')}{\partial p} &=& 0 
\end {array}
\right. \ \mbox { for } \ \eta < 0 
 \, . 
\ee
 For instance, in the case of focusing nonlinearity then we have that the first two bifurcation points for $\alpha =-4$ and $a=1$ are given by
\begin {itemize}

\item [-] $(\eta_1 , \Omega_1 )=(-19.35(4),-6.82(5))$;

\item [-] $(\eta_2 , \Omega_2 )= (-81.74(0),-54.41(7)))$.

\end {itemize}

\begin {remark} \label {Nota8}
Let $\alpha <0$ and let $\varphi := \varphi_\Omega $ be the stationary ground state solution to (\ref {Eq38}) with norm $\mu = \| \varphi_\Omega \|_{L^2}$. \ In order to establish if such a solution is stable or not we make use of the following criterion (see Proposition 3 by \cite {FOO} where $\omega =-\Omega$, see also \cite {FO1,FO2,F}) for $\eta <0$. \  Assuming that some spectral assumptions are satisfied then the crucial point to check is the slope condition; that is, if: 
\begin {itemize}

 \item [-] $\frac {\partial \mu^2}{\partial \omega }
 >0$ then the solution is stable;
 
 \item [-] $\frac {\partial \mu^2}{\partial \omega }<0$ then the solution is unstable.
\end {itemize}
Hence, from the numerical experiment (see Fig. \ref {Fig1}) we can conjecture that the stationary solution associated to $\En_0 (\eta )$ is stable.
\end {remark}

\subsection {Stationary states for defocusing NLS with cubic nonlinearity} \label {S4_3}

In this section we restrict our attention to the defocusing nonlinearity case where $g=+1$, thus the equation takes the form 
\bee
-\frac {d^2 \varphi}{dx^2} + \varphi^3 = {\En} \varphi \,  , \label {Eq41}
\eee
and, in this case, it is more convenient to write the general solution in the form
\be
\varphi (x) = C \cs \left ( \lambda (x-x_0 ) , p \right ) \, , \ p \in [0,1]\, , 
\ee
where $\cs (u,p)= \frac {\cn(u,p)}{\sn(u,p)}$ and 
\be
\En = \lambda^2 (p^2-2) \ \mbox { and } \ 
C^2 = 2\lambda^2 \, ,
\ee
for some $C,\lambda \in \R$. \ Recalling that 
\be
\lim_{p \to 1^-} \cs(u,p) = \ch (u)= \frac {1}{\mbox {sinh}(u)}
\ee
and since we look for a real-valued solution $\varphi (x)$  to (\ref {Eq41}) such that $\varphi (x)=0$, for $x\le 0$, and $\varphi (x) \to 0$ as $x\to + \infty$ then such a solution there exists only when ${\En}<0$ and it has the form 
\be
\varphi (x) = 
\left \{
\begin {array}{ll}
C \cs \left ( \lambda (x-x_0 ) , p \right ) & ,\ x \in (0,a) \ \mbox { where }\En = \lambda^2 (p^2-2) \ \ \mbox { and } \ C^2 = 2\lambda^2  \\ 
C' {\ch} \left ( \lambda' (x-x_0' ) \right ) & ,\ a< x \ \mbox { where } {\lambda'}^2 =-{\En} \ \mbox { and } \ {C'}^2 =  - {2{\En}} 
\end {array}
\right. \, .
\ee
Also in this case we may assume, for argument's sake and without loosing in generality, that $\lambda >0$ and $\lambda ' >0$ (if not one can change the sign of $C$ and $C'$).

The Dirichlet boundary condition $\varphi (0)=0$ at $x=0$ implies that $\lambda x_0 ={\KEll}(p)$ is a zero of the Jacobi elliptic function $\cn (x,p)$. \ The matching condition (\ref {Eq4}) at $x=a$ implies that 
for $g=+1$ the real-valued parameters $C$, $C'$, $\lambda$, $\lambda'$, $ x_0$, $ x_0'$  and $ p$ must satisfy to the following set of conditions
\bee
\left \{
\begin {array}{l}
\lambda x_0 ={\KEll}(p)\\
C' {\ch} \left ( \lambda' (a-x_0' ) \right ) - C \cs \left ( \lambda (a-x_0 ) , p \right ) =0 \\
C' {\ch} (\lambda' (a-x_0')) \left [ \lambda' \mbox {coth} (\lambda' (a-x_0')) + \alpha \right ] - C \lambda \frac {\mbox {dn} \left ( \lambda (a-x_0 ) , p \right )}{ {\sn}^2 \left ( \lambda (a-x_0 ) , p \right )}
= 0\\
 {C}^2 = {2{\lambda}^2} \\
 {C'}^2 = {2{\lambda'}^2} \\
 -(2-p^2) \lambda^2 = -{\lambda '}^2 =\Omega
\end {array}
\right. \label {Eq42}
\eee
with the constrain $p \in [0,1]$. 

Since $\mbox {\rm coth} (x) =\pm \sqrt {1+{\ch}^2 (x)}$ then by reduction we can reduce the system of equations as follows (also the proof of this result is just a straightforward calculation we omit).

\begin {lemma} [Effective equation for stationary solutions - repulsive nonlinearity] \label {Lemma4} 
Let 
\be
u(p):=\frac {1}{\sqrt {2-p^2}}
\ee
and let 
\be
H_\ell^d (p,\lambda ') &:=&  \cn \left [ u(p) \lambda' a-{\KEll}(p)  , p \right ] \left \{ \lambda'  (-1)^\ell \sqrt {1 + u^2(p) {\cs}^2 \left [u(p) 
\lambda' a-{\KEll}(p),p \right ] } + \alpha \right \} + \\ 
&& \ \ -  u(p) \lambda' \frac {\mbox {\rm dn} \left [ u(p) \lambda' a-{\KEll}(p) , p \right ]}{{\sn} \left [ u(p) \lambda' a-{\KEll}(p) , p \right ]} \, , \ \ell=1,2 \, .
\ee
If $p$, $x_0$, $x_0'$, $\lambda$, $\lambda'$ $C$ and $C'$ is a solution to (\ref {Eq42}) then the following equation must be satisfied
\bee
H_\ell^d (p,\lambda' )=0 \, ,\label {Eq43}
\eee 
for $\ell =1$ or $\ell =2$. \ Furthermore, the norm of the solution is given by:
\be
\mu^2 = \| \varphi \|^2 = C^2 \int_0^a \left [ {\cs} \left ( \lambda (x-x_0 ) , p \right ) \right ]^2 dx + {C'}^2 \int_a^{+\infty} \left [ {\ch} \left ( \lambda' (x-x_0' ) \right ) \right ]^2 dx  
\ee
\end {lemma}

\appendix 

\section {Conservation of the norm and of the energy, calculation of the derivatives of the variance.} \label {AppA}

By means of standard regularization arguments as done by \cite {ASa} it follows that the solution $\psi_t$ is such that $x\psi_t \in L^2$, for any $t \in [0,T_{max})$, and we can simply consider a formal calculation of the variance $I_q$ and of its derivatives $\dot I_q $ and $\ddot I_q := \frac {d^2 I_q}{dt^2}$, as well as of norm of $\psi_t$ and of the energy. \ To this end we remind that $\psi_t$  satisfies the boundary conditions (\ref {Eq3}) and (\ref {Eq4}) and that $\psi_t (x) =0 $ for any $x \le 0$; in particular we have that $\dot \psi_t (0)=0$ since $\psi_t (0)=0$ for any $t$.

Conservation of the norm is trivial. 

In order to check the conservation of energy let 
\be
h(t) = \| \psi_t \|_{L^{2\sigma +2}}^{2\sigma +2} \ \mbox { and } \  q(t) = \langle \psi_t , H \psi_t \rangle 
\ee
and so 
\be
{\mathcal E} (\psi _t )= q(t)+\frac {\eta}{\sigma +1} h(t)\, .
\ee
We note that
\be
\dot h &=& \frac {d}{dt}  \langle \psi_t^{\sigma +1} , \psi_t^{\sigma +1} \rangle =
(\sigma +1 ) \left [ \langle \dot \psi_t , |\psi_t |^{2\sigma }\psi_t \rangle +  \langle |\psi_t |^{2\sigma } \psi_t , \dot \psi_t \rangle \right ] \\
&=&
2(\sigma +1 ) \Re \langle \dot \psi_t , |\psi_t |^{2\sigma }\psi_t \rangle = 2 (\sigma +1) \Re \left \{ i \langle H \psi_t + \eta |\psi_t |^{2\sigma } \psi_t , |\psi_t|^{2\sigma } \psi_t \rangle \right \} \\
&=& - 2 (\sigma +1) \Im  \langle- \psi_t'' , |\psi_t |^{2\sigma }\psi_t \rangle \, , \
 \ee
and (here $'$ denotes the derivative with respect to $x$)
\be
\dot q(t) &=& \langle \dot \psi_t , H \psi_t \rangle + \langle \psi_t , H \dot \psi_t \rangle = 2 \Re \langle \dot \psi_t , H \psi_t \rangle \\
&=& 2 \Re i\langle i\dot \psi_t , H \psi_t \rangle = -2 \Im \left [ \langle H \psi_t , H \psi_t \rangle  + \langle \eta |\psi_t |^{2\sigma } \psi_t , H \psi_t \rangle \right ]  \\
&=& - 2 \eta \Im \langle \psi_t'' , |\psi_t |^{2\sigma } \psi_t \rangle 
\ee
Hence 
\be
\dot q + \frac {\eta}{\sigma +1} \dot h =0 \, . 
\ee

Concerning the variance $I_q = \langle \psi_t, (x-q)^2 \psi_t \rangle$ we remark that $(x-q)^2 \psi_t$ satisfies the boundary conditions  (\ref {Eq3}) and (\ref {Eq4}), then we have that
\be
\dot I_q 
&=& i \langle i \dot \psi_t , (x-q)^2 \psi_t \rangle - i  \langle \psi_t , (x-q)^2 i\dot \psi_t \rangle  \\ 
&=&i \langle  \psi_t ,  H(x-q)^2 \psi_t \rangle - i  \langle \psi_t , (x-q)^2 H \psi_t \rangle     \\ 
&=& -i \langle  \psi_t , [2 \psi_t + 4 (x-q) \psi_t ' +(x-q)^2 \psi_t'']   \rangle + i  \langle (x-q)^2  \psi_t , \psi_t'' \rangle \\ 
&=& -2i \| \psi_t \|^2 -4i  \langle  \psi_t ,(x-q) 
\psi_t '  \rangle \, . 
\ee
Thus (\ref {Eq31}) and (\ref {Eq32}) hold true since $I_q$ is real-valued. \ For what concerns the second derivative of the variance  a straightforward calculation gives that (let $K=H+\eta |\psi|^{2\sigma }$)
\be
\ddot I_q &=& \frac {d}{dt} 4 \Im \langle \psi_t , (x-q) \psi_t' \rangle \\ 
&=& 4 \Im \langle  \dot \psi_t , (x-q) \psi_t ' \rangle +  4 \Im \langle  \psi_t , (x-q)  \dot \psi_t ' \rangle 
\\
&=& 4  \Re  \langle K \psi_t ,  \psi_t  \rangle +  8 \Re \langle  (x-q) \psi_t' ,  K \psi_t \rangle \\
&=& -  8 \Re \langle  (x-q) \psi_t' ,  \psi_t'' \rangle  +  8 \eta \Re \langle  (x-q) \psi_t' ,|\psi_t|^{2\sigma} \psi_t \rangle  + 4  \Re  \langle K \psi_t ,  \psi_t  \rangle  \\
&=& -  4 \Re \langle  (x-q) \psi_t' ,  \psi_t'' \rangle   -  4 \Re \langle  (x-q) \psi_t' ,  \psi_t'' \rangle +  8 \eta \Re \langle  (x-q) \psi_t' ,|\psi_t|^{2\sigma} \psi_t \rangle  + 4  \Re  \langle K \psi_t ,  \psi_t  \rangle
\ee

We remark that
\be
\langle  (x-q) \psi_t' ,  \psi_t'' \rangle = 
-\langle \psi_t',\psi_t' \rangle - \langle (x-q) \psi_t'',\psi_t' \rangle + \T
\ee
where we set $\T= q | \psi_t ' (0+)|^2  -(a-q) \left [| \psi_t ' (a+) |^2 - | \psi_t ' (a-) |^2 \right ]$. \ Hence,
\be
\ddot I_q 
&=&  4 \Re \langle \psi_t ', \psi_t ' \rangle + 4 \Re \langle (x-q) \psi_t '', \psi_t ' \rangle - 4  \Re  \T -  4 \Re \langle  (x-q) \psi_t' ,  \psi_t'' \rangle + \\
&& \ \ + 8 \eta \Re \langle  (x-q) \psi_t' ,|\psi_t|^{2\sigma} \psi_t \rangle  +  4  \Re  \langle K \psi_t ,  \psi_t  \rangle \\
 &=&  4 \| \psi_t '\|^2  - 4   \T
 + 8 \eta \Re \langle  (x-q) \psi_t' ,|\psi_t|^{2\sigma} \psi_t \rangle  +  4  \Re  \langle H \psi_t ,  \psi_t  \rangle +  4  \Re  \langle \eta |\psi_t|^{2\sigma} \psi_t ,  \psi_t  \rangle \\
 &=&  8 \| \psi_t '\|^2  - 4   \T
 + 8 \eta \Re \langle  (x-q) \psi_t' ,|\psi_t|^{2\sigma} \psi_t \rangle  +  4  \alpha |\psi (a)|^2 +  4 \eta \| \psi_t\|^{2\sigma +2}_{L^{2\sigma +2}}   
\ee
Recalling that
\be 
{\mathcal E} (\psi_t )= \| \psi_t ' \|^2 + \alpha | \psi_t (a) |^2 +  \frac 1{\sigma +1} \eta \| \psi_t \|_{L^{2(\sigma +1)}}^{2(\sigma +1)} = const. := {\mathcal E}_0 
\ee
from which follows that
\be
 \| \psi_t ' \|^2 = {\mathcal E}_0  - \alpha | \psi_t (a) |^2 - \frac 1{\sigma +1} \eta \| \psi_t \|_{L^{2(\sigma +1)}}^{2(\sigma +1)}\, . 
 \ee
Hence
 \be
 \ddot I_q &=& 8 \left [ {\mathcal E}_0  - \alpha | \psi_t (a) |^2 - \frac 1{\sigma +1} \eta \| \psi_t \|_{L^{2(\sigma +1)}}^{2(\sigma +1)} \right ] + \\ 
 && \ - 4   \T
 + 8 \eta \Re \langle  (x-q) \psi_t' ,|\psi_t|^{2\sigma} \psi_t \rangle  +  4  \alpha |\psi (a)|^2 +  4 \eta \| \psi_t\|^{2(\sigma +1)}_{L^{2(\sigma +1)}}   \\
&=& 8  {\mathcal E}_0  - 4 \alpha | \psi_t (a) |^2 + \frac {4\eta (\sigma -1)}{\sigma +1} \| \psi_t \|_{L^{2(\sigma +1)}}^{2(\sigma +1)} - 4 \T     +   8 \eta \Re \langle  (x-q) \psi_t' , |\psi_t |^{2\sigma}  \psi_t \rangle \, . 
\ee
Finally
\be
\langle  (x-q) \psi_t' , |\psi_t |^{2\sigma}  \psi_t \rangle 
&=&- \langle   \psi_t , \bar \psi_t^\sigma \psi_t^{\sigma +1} \rangle - \sigma \langle   \psi_t , (x-q) \bar \psi_t^{\sigma -1} \bar \psi_t '  \psi_t^{\sigma +1} \rangle - (\sigma +1)  \langle   \psi_t ,  (x-q) \bar \psi_t^\sigma \psi_t^{\sigma} \psi_t' \rangle \\ 
&=& - \| \psi_t \|_{L^{2(\sigma +1)}}^{2(\sigma +1)}  - \sigma \langle   \psi_t' , (x-q) | \psi_t|^{2\sigma}  \psi_t \rangle - (\sigma +1)  \langle   (x-q) |\psi_t|^{2\sigma} \psi_t ,   \psi_t' \rangle 
\ee
from which follows that
\be
\Re \langle  (x-q) \psi_t' , |\psi_t |^{2\sigma}  \psi_t \rangle = - \frac {1}{2(\sigma +1)} \| \psi_t \|_{L^{2(\sigma +1)}}^{2(\sigma +1)} \, . 
\ee
Eventually, we have that
\be
\ddot I_q 
 &=&  8  {\mathcal E}_0  - 4 \alpha | \psi_t (a) |^2 + \frac {4\eta (\sigma -1)}{\sigma +1} \| \psi_t \|_{L^{2(\sigma +1)}}^{2(\sigma +1)} - 4 \T     -   4 \eta \frac {1}{(\sigma +1)} \| \psi_t \|_{L^{2(\sigma +1)}}^{2(\sigma +1)} \\ 
 &=&  8  {\mathcal E}_0  - 4 \alpha | \psi_t (a) |^2 + \frac {4\eta (\sigma -2)}{\sigma +1} \| \psi_t \|_{L^{2(\sigma +1)}}^{2(\sigma +1)}  \\ 
 &=&  8  {\mathcal E}_0  - 4 \alpha | \psi_t (a) |^2 + \frac {4\eta (\sigma -2)}{\sigma +1} \| \psi_t \|_{L^{2(\sigma +1)}}^{2(\sigma +1)} -4\T 
\ee
completely so the formal calculations.

\end {document}